\documentclass[11pt]{article}
\usepackage[latin1]{inputenc} 
\usepackage{amssymb}
\usepackage{amsfonts} 
\usepackage{graphicx} 
\usepackage{amsthm} 
\usepackage{amsmath} 
\usepackage{epsfig}

\def\nn{\nonumber}

\def \bc {\begin{center}}
\def \ec {\end{center}}
\def \bi {\begin{itemize}}
\def \ei {\end{itemize}}

\def \ba {\begin{array}}
\def \ea {\end{array}}

\def \bea {\begin{eqnarray}}
\def \eea {\end{eqnarray}}

\def \be {\begin{equation}}
\def \ee {\end{equation}}

\def \ce {E}

\def\cG {{\cal G}}

\def\w#1{{v}_#1}
\def\z#1{{x}_#1}
\def\u#1{{u}_#1}

\theoremstyle{remark}

\newcommand{\la}{\langle}
\newcommand{\ra}{\rangle}

\begin{document}

\begin{center}
{\LARGE {\bf Thermal Vacuum Radiation in Spontaneously Broken Second-Quantized
Theories on Curved Phase Spaces of Constant Curvature}}
\end{center}
\bigskip
\bigskip

\centerline{{\sc M. Calixto}$^{1,2}$ {\sc and} {\sc V. Aldaya}$^{2}$}

\bigskip

\bc {\it $^1$ Departamento de Matemática Aplicada y Estad\'\i stica,
Universidad Politécnica de Cartagena, Paseo Alfonso XIII 56, 30203
Cartagena, Spain}
\\
{\it $^2$ Instituto de Astrof\'\i sica de Andaluc\'\i a (IAA-CSIC),
Apartado Postal 3004, 18080 Granada, Spain}

\bigskip

 Manuel.Calixto@upct.es \hskip 2cm valdaya@iaa.es\ec

\bigskip

\bigskip
\begin{center}
{\bf Abstract}
\end{center}
\small
\begin{list}{}{\setlength{\leftmargin}{3pc}\setlength{\rightmargin}{3pc}}
\item
We construct second-quantized (field) theories on coset spaces of
pseudo-unitary groups $U(N_+,N_-)$. The existence of degenerated quantum
vacua (coherent states of zero modes) leads to a breakdown of the original
pseudo-unitary symmetry. The action of some spontaneously broken symmetry
transformations destabilize the vacuum and make it to \emph{radiate}. We
study the structure of this \emph{thermal} radiation for curved phase
spaces of constant curvature: complex projective spaces
$\mathbb CP^{N-1}=SU(N)/U(N-1)$ and open complex balls $\mathbb
CD^{N-1}=SU(1,N-1)/U(N-1)$. Positive curvature is related to generalized
Fermi-Dirac (FD) statistics, whereas negative curvature is connected with
generalized Bose-Einstein (BE) statistics, the standard cases being
recovered for $N=2$. We also make some comments on the contribution of the
vacuum (dark) energy to the cosmological constant and the phenomenon of
inflation.
\end{list}
\normalsize 




\section{Introduction}
Quantum vacua are not really empty to every observer. Actually, the
quantum vacuum is filled with zero-point fluctuations of quantum fields.
The situation is similar to quantum many-body condensed mater systems
describing, for example, superfluidity and superconductivity, where the
ground state mimics the quantum vacuum in many respects and quasiparticles
(particle-like excitations above the ground state) play the role of
matter.

We know that zero-point energy, like other non-zero vacuum expectation
values, leads to observable consequences as, for instance, the Casimir
effect, and influences the behaviour of the Universe at cosmological
scales, where the vacuum (dark) energy is expected to contribute to the
cosmological constant, which affects the expansion of the universe, i.e.
the phenomenon of inflation (see e.g. \cite{vacmith} for a nice review).
Indeed, dark energy is the most popular way to explain recent observations
that the universe appears to be expanding at an accelerating rate.

Also, in Quantum Field Theory, the vacuum state is expected to be stable
under some underlying group of symmetry transformations $G$ (namely, the
Poincaré group). Then the action of some spontaneously broken symmetry
transformations can destabilize/excitate the vacuum and make it to
\emph{radiate}. Such is the case of the Planckian radiation of the
Poincaré invariant vacuum under uniform accelerations, that is, the
Fulling-Unruh effect \cite{Fulling,Unruh}, which shares some features with
the (black-hole evaporation) Hawking effect \cite{Hawking}. Here, the
Poincaré invariant vacuum looks the same to every inertial observer but
converts itself into a thermal bath of radiation with temperature $T=\hbar
a/2\pi ck_B$ in passing to a uniformly accelerated frame ($a$ denotes the
acceleration,
$c$ the speed of light and $k_B$ the Boltzmann constant). These radiation
phenomena are usually linked to some kind of global mutilation of the
spacetime (viz, existence of horizons). In the reference \cite{conforme},
it was shown that the reason for this radiation is more profound and
related to the spontaneous breakdown of the conformal symmetry in
\emph{quantum} field theory. From this point of view, a Poincaré invariant
vacuum can be regarded as a coherent state of conformal zero modes, which
are undetectable (`dark') by inertial observers but unstable under
relativistic uniform accelerations (special conformal transformations).
There we used the conformal group in (1+1)-dimensions,
$SO(2,2)\simeq SO(2,1)\times SO(2,1)$, which consists of two copies of the
pseudo-orthogonal group $SO(2,1)$ (left- and right-moving modes,
respectively).

In \cite{jpa} we constructed $O(3), O(2,1)$ and Newton-Hooke invariant
quantum field theories in a unified manner. We could think of
$O(3)$ and
$O(2,1)$ as isometry subgroups of the spatial part of de Sitter and Anti de Sitter spaces,
with positive and negative curvature $\kappa$, respectively. In
\cite{brokenFD} we studied the thermodynamics of the excited ground state
of a quantum many-body system with broken unitary
$U(N)$ symmetry and their relation to Fermi-Dirac statistics,
restricting ourselves to the fundamental representation.

In this paper we extend this construction to general (pseudo)-unitary
symmetry groups. We study the spontaneous breakdown of the pseudo-unitary
symmetry in second-quantized field theories on coset spaces of
$U(N_+,N_-)$. Specially, we shall concentrate on curved phase spaces of
constant curvature: complex projective spaces
$\mathbb CP^{N-1}=SU(N)/U(N-1)$ and open complex balls $\mathbb
CD^{N-1}=SU(1,N-1)/U(N-1)$, with positive and negative curvature,
respectively.

We shall show that, similar to the \emph{spin-statistics theorem} in
quantum field theory relating the spin of a particle to the statistics
obeyed by that particle, a sort of \emph{curvature-statistics connection}
can also be established in our context. In fact, a degenerated ground
state (actually, a coherent state of zero modes) radiates as a black body
for second-quantized field theories on curved phase spaces of negative
curvature, while the quantum statistics of the radiation is of Fermi-Dirac
(FD) type for positive curvature.

The organization of the paper is as follows. In Sec. \ref{generalback} we
describe the general construction of a $G$-invariant quantum many-body
system as a second quantization on a group $G$ and the spontaneous
breakdown of
$G$ to $G_0$, the stability subgroup of the degenerated vacuum. In Sec.
\ref{structureofpseudo} we discuss quantum mechanics on coset spaces of
pseudo-unitary groups $U(N_+,N_-)$ and its relation to non-linear
sigma-models. We introduce a parametrization of general pseudo-unitary
groups and describe their coset spaces and unitary irreducible
representations, using the machinery of Coherent States
\cite{Perelomov,Klauder}. They are the ingredients to calculate
probability distributions, average energies and the thermodynamics of the
excited ground state in Sec. \ref{genqstat}, whose statistics turns out to
generalize the usual FD and BE expressions, depending on the curvature.
The last Section is devoted to some comments on the vacuum (\emph{dark})
energy and its implications in cosmology.

\section{The general context}\label{generalback}

Let ${\cal G}=\{X_\alpha, \,\alpha=0,1,2,\dots,l\}$ be the (Lie) algebra
of observables of a given quantum system, among which we highlight $X_0$
as the Hamiltonian operator. Let ${\cal H}$ be the (Hilbert) carrier space
of a unitary irreducible representation $U$ of the Lie group $G$. Let us
assume that the energy spectrum is discrete and bounded from bellow, that
is, there is a vacuum vector $|0\rangle$ whose energy $E_0$ we set to
zero, i.e.
\be
X_0|0\rangle=0. \label{requirement}\ee
Let $B({\cal H})=\{|n\rangle,
\,n=0,1,2,\dots\}$ be a (finite or infinite) orthonormal basis of ${\cal
H}$ made up of energy eigenvectors, $X_0|n\rangle=E_n|n\rangle$.

An important ingredient to construct a $G$-invariant quantum field theory,
as a second-quantized (many-particle) theory, will be the irreducible
matrix coefficients
\be U_{mn}(x)\equiv \la m|U(x)|n\ra, \label{imc}\ee
of the representation $U(x)=e^{i \sum_\alpha x^\alpha X_\alpha}$ of $G$ in
the orthonormal basis $B({\cal H})$, where $\{x^\alpha\in\mathbb{R},
\alpha=0,\dots,{\rm dim}(G)-1\}$ stands for a coordinate system in $G$.

Given the Fourier expansion, in energy modes, of a state
\be
|\psi\rangle=\sum_{n=0} a_n|n\rangle\,,\label{campoads} \ee
the Fourier coefficients $a_n$ (resp. $a^\dag_n$) are promoted to
annihilation (resp. creation) operators of energy modes $E_n$ in the
second quantized theory, with commutation relations
$[a_n,a^\dag_m]=\delta_{n,m}$.  The (finite) action of $G$ on annihilation operators is:
\be
a_m\rightarrow a'_m=\sum_{n=0} U_{mn}(x) a_n, \label{transf1}\ee
together with the conjugated expression for creation operators. The
infinitesimal generators of this finite action are the second quantized
version, $\hat{X}_\alpha$, of the basic observables $X_\alpha$. They have
the following explicit expression in terms of creation and annihilation
operators:
\be
\hat{X}_\alpha=-i\sum_{m,n=0} a^\dag_m\left.\frac{\partial
U_{mn}(x)}{\partial x^\alpha}\right|_{x=0} a_n.\label{observables2}\ee
For example, since $|n\ra$ are energy $E_n$ eigenstates and we have set
$E_0=0$, the Hamiltonian operator is:
\be \hat{X}_0=\sum_{n=1} E_n a^\dag_n a_n.\label{energy2} \ee

The vacuum $|\hat{0}\rangle$ of the second quantized theory is
characterized by being stable under the symmetry group $G$, i.e. it is
annihilated by the symmetry generators
\be \hat{X}_\alpha|\hat{0}\rangle=0,\,\, \alpha=0,\dots,{\rm dim}(G)-1\ee
and also by $a_n$, i.e. $a_n|0\rangle=0, \forall n\geq 0$. Then an
orthonormal basis for the Hilbert space of the second quantized theory is
constructed by taking the orbit through the vacuum $|\hat{0}\rangle$ of
the creation operators
$a^\dag_n$:
\be
|q(n_1), \dots, q(n_p)\rangle\equiv\frac{(a^\dag_{n_1})^{q(n_1)}\dots
(a^\dag_{n_p})^{q(n_p)} }{(q(n_1)!\dots
q(n_p)!)^{1/2}}|\hat{0}\rangle,\label{orbit1} \ee
where $q(n)\in \mathbb N$ denotes the occupation number of the energy
level $n$.

Note that any multi-particle state (\ref{orbit1}) made up of an arbitrary
content of \emph{zero modes}, like
\be
|\theta\rangle\equiv\sum_{q=0}^\infty
\theta_q(a^\dag_0)^q|\hat{0}\rangle,\;\;\theta_q\in
\mathbb{C},\label{thetav} \ee
has zero total energy, i.e. $\hat{X}_0|\theta\rangle=0$, since
$[\hat{X}_0,a_0]=0$. It also verifies
$a_n|\theta\rangle=0, \forall n>0$. Let us denote by
$G_0\subset G$ the maximal stability (isotropy) subgroup of this degenerated ground state
$|\theta\ra$, of which the Hamiltonian $\hat{X}_0$ is one of its
generators.

Actually, the operator $a_0$ conmutes with the unbroken symmetry
generators
$\hat{X}_\alpha^{(0)}\in {\cal G}_0$ (the Lie algebra of $G_0$) and the creation operators:
\be
[a_0,\hat{X}_\alpha^{(0)}]=0=[a_0,a^\dag_n],\; \forall n>0,
\label{unbrokengen}\ee
so that, by Schur's Lemma, $a_0$ behaves as (a multiple of) the identity
operator in the broken theory. That is, it is natural to demand
$a_0$ to leave the $G_0$-invariant vacuum (\ref{thetav}) stable, which implies
that:
\be
a_0|\theta \rangle= \theta |\theta\rangle\Rightarrow
|\theta\rangle=e^{\theta a^\dag_0-\bar{\theta}a_0}|\hat{0}\rangle.
\label{thetav2}\ee
Thus, the vacuum of our (spontaneously) broken theory will be a
\emph{coherent state of zero modes} (see \cite{Perelomov} and
\cite{Klauder} for a thorough exposition on coherent states).

Now we show that a general unitary symmetry transformation
(\ref{transf1}), which incorporates broken symmetry generators in
$G/G_0$, produces a ``rearrangement'' of this  pseudo-vacuum
$|\theta\rangle$ and causes it to radiate. In other words, we can associate a
\emph{quantum statistical ensemble} with the excited (or, let us say,
``polarized'') vacuum
\be |\theta'\rangle\equiv e^{\theta
a'^\dag_0-\bar{\theta}a'_0}|\hat{0}\rangle, \label{vacp} \ee
as follows.

The average number of particles with energy $E_n$ can be computed as the
expectation value:
\be
 N_n(x)=\la\theta'|a^\dag_na_n|\theta'\ra=|\theta|^2 |U_{0n}(x)|^2, \label{averagenp}\ee
hence, $|\theta|^2$ is the total average of particles of this quantum
statistical ensemble. In the same way, the probability $P_n(q,x)$ of
observing $q$ particles with energy $E_n$ in
$|\theta'\ra$ can be calculated as:
\be
P_n(q,x)=|\la q(n) |\theta'\ra|^2=\frac{e^{-|\theta|^2}}{q!}|\theta|^{2q}
\, |U_{0n}(x)|^{2q}=\frac{e^{-|\theta|^2}}{q!}N_n^{q}(x).\label{probpart}
\ee
Therefore, the relative probability of observing a state with total energy
$E$ in the excited vacuum
$|\theta'\rangle$ is:
\be P(E)=\sum_{\begin{array}{c} q_0,\dots,q_k :\\ \sum^k_{n=0}E_n
q_n=E\end{array}} \prod^k_{n=0}P_n({q}_n,x)\,.\label{probest}\ee
For the cases studied in this paper, this distribution function can be
factorized as
$P(E)=\Omega(E)e^{-\tau E}$, where
$\Omega(E)$ is a relative weight proportional to the number of states with
energy $E$ and the factor
$e^{-\tau E}$ fits this weight properly to a temperature $T=k_B/\tau$.

We shall be primarily interested in the mean values of the basic
observables (\ref{observables2}). They can be calculated as:
\bea {\cal X}_\alpha&\equiv&\la \theta'|\hat{X}_\alpha|\theta'\ra=
-i|\theta|^2\sum_{m,n=0} U_{0m}(x)\frac{\partial U_{mn}(0)}{\partial
x^\alpha}\bar{U}_{0n}(x)\nn\\ &=&-i|\theta|^2 \left(U(x)\frac{\partial
U(0)}{\partial x^\alpha}U^\dag(x)\right)_{00}.\label{meanobservables}\eea
In particular, the mean energy is simply:
\be
{\cal X}_0=|\theta|^2 \sum_{n=1} |U_{0n}(x)|^2 E_n.\label{meanenergy}\ee
We shall see that ${\cal X}_0$ generalizes the usual FD and BE expressions
for the compact $G=U(N)$ and non-compact $G=U(1,N-1)$ cases, respectively.

\section{Quantum mechanics on cosets of $U(N_+,N_-)$}\label{structureofpseudo}

We shall focus on (pseudo-)unitary groups:
\be G=U(N_+,N_-)=\{U\in M_{N\times N}(\mathbb{C})\,\,/\,\,U\Lambda
U^\dag=U^\dag \Lambda U=\Lambda,\,\,N=N_++N_-\}, \ee
i.e., groups of complex $N\times N$ matrices $U$ that leave invariant the
indefinite metric \[\Lambda={\rm
diag}(1,\stackrel{N_+}{\dots},1,-1,\stackrel{N_-}{\dots},-1).\] As the
Lie-algebra $\cG$ generators we can choose the step operators
${X}_{\alpha}^{\beta}$,
\be {\cal G}=u(N_+,N_-)=\langle{X}_{\alpha}^{\beta},\;\;{\rm with}\;\;
({X}_{\alpha}^{\beta})_{\mu}^{\nu}\equiv
\delta_{\alpha}^{\nu}\delta^{\beta}_{\mu};\,\alpha,\beta,\mu,\nu=1,\dots,N\rangle,\label{pun}\ee
fulfilling the commutation relations:
\be \left[{{X}}_{\alpha_1}^{\beta_1},{{X}}_{\alpha_2}^{\beta_2}\right]=
\delta_{\alpha_1}^{\beta_2}{{X}}_{\alpha_2}^{\beta_1} -
\delta_{\alpha_2}^{\beta_1}{{X}}_{\alpha_1}^{\beta_2}, \label{bosreal}\ee
and the usual orthogonallity relations:
\be
{\rm
Tr}(X_{\alpha}^{\beta}X_{\gamma}^{\rho})=\delta_{\alpha}^{\rho}\delta_{\gamma}^{\beta}.
\ee
The Cartan (maximal Abelian) subalgebra ${\cal C}=u(1)^N$ is made of
diagonal operators
\be
{\cal C}=\langle X_{\alpha}^{\alpha},
\alpha=1,\dots,N\rangle.\label{cartan}\ee
We shall choose our Hamiltonian
$X_0$ to be a linear combination of Cartan generators
\be X_0=\sum_{\alpha=1}^{N} \lambda_\alpha
X_{\alpha}^{\alpha},\label{hamiltonian}\ee
and the Hilbert space
${\cal H}$ to be a lowest-weight unitary irreducible representation of $G$.
For unitary irreducible representations of $G$ we have the conjugation
relation (summation over repeated indices will be used where no confusion
arises):
\be
X_{\alpha\beta}^\dag=X_{\beta\alpha},\;\;X_{\alpha\beta}\equiv
\Lambda_{\alpha\gamma} X^\gamma_\beta,\ee
which distinguishes between compact and non-compact cases
(positive-definite and indefinite metrics $\Lambda$, respectively).

\subsection{Non-linear sigma-models on cosets of $U(N_+,N_-)$}

For those readers who prefer the Lagrangian picture to a
group-theoretic/algebraic viewpoint, we shall consider, for example, the
case of $G$-invariant non-linear sigma-models described by the
\emph{partial-trace} Lagrangian
\bea {\cal L}(U,\dot U)&=&\rho_1{\rm Tr}_{G^0} (\vartheta^L)+\rho_2{\rm
Tr}_{G/G^0} (\vartheta^L)^2\equiv \rho_1{\rm Tr}
(X_0\vartheta^L)+\rho_2{\rm Tr} ([iX_0,\vartheta^L])^2\nn\\
&=&\rho_1\sum_{\alpha=1}^{N}\lambda_\alpha \vartheta^\alpha_\alpha +\rho_2
\sum_{\alpha,\beta=1}^{N}(\lambda_\alpha-\lambda_\beta)^2\vartheta^\alpha_\beta\vartheta^\beta_\alpha,
\label{Lagrangian}\eea
where
\be
\vartheta^L= iU^{-1}\dot U=\vartheta_\alpha^\beta X^\alpha_\beta,\,\,\,
\vartheta^\alpha_\beta=\bar{\vartheta}_\alpha^\beta \ee
stands for the canonical left-invariant 1-form on
$G$ and $\rho_{1,2}$ are coupling constants. This Lagrangian is obviously left-invariant under rigid (global)
transformations
\be
U\to U'U, \ \forall U'\in G \label{leftrans}\ee
and right-invariant under local (gauge) transformations
\be
U\rightarrow UU_0(t),\ U_0(t)\in G^0,\label{gauge}\ee
belonging to the isotropy subgroup $G^0\subset G$ of $X_0$ under the
adjoint action $X_0\rightarrow UX_0U^{-1}$ of the group $G$ on its Lie
algebra ${\cal G}$. In general, we shall have
$G^0\subset G_0$, where $G_0$ is the maximal stability (isotropy) subgroup
of the degenerated vacuum $|\theta\rangle$, whose generators fulfil
(\ref{unbrokengen}).

Let us see some examples for
$G=U(N)$:
\begin{enumerate}
\item  For the case $\lambda_\alpha\not=\lambda_\beta,\forall \alpha,\beta=1\dots,N$, the unbroken symmetry group is
$G^0=U(1)^{N}$ and the Lagrangian (\ref{Lagrangian}) describes a motion on the coset (\emph{flag
manifold}) $\mathbb F_{N-1}=G/G^0=U(N)/U(1)^{N}$.
\item For $\lambda_\alpha=0, \forall \alpha>1$,
the gauge group is
$G^0=U(N-1)\times U(1)$ and  (\ref{Lagrangian}) describes a motion on the \emph{complex projective} space
$\mathbb CP^{N-1}=U(N)/U(N-1)\times U(1)$. Eventually, we shall restrict ourselves to this case and set,
for example,
\be X_0=\lambda X^1_1,\label{hamiltoniancpn}\ee
up to constant (zero-point energy) addends --see later on Eqs.
(\ref{hamiltonianu2}) and (\ref{higherhamilcpn1-1}).
\item For other choices like:
\[0=\lambda_1=\lambda_2=\dots=\lambda_{N_1}\not=\lambda_{N_1+1}\not=\dots\not=\lambda_{N-N_2}=\dots=\lambda_{N}=0\]
the unbroken symmetry group is $G^0=U(N_1)\times U(N_2)\times U(1)$ and
(\ref{Lagrangian}) describes a motion on the \emph{complex Grasmannian}
manifold $\mathbb CG(N_1,N_2)=U(N)/U(N_1)\times U(N_2)\times U(1)$ (see
next Section for suitable coordinate systems on these coset spaces).
\end{enumerate}

At the quantum level, gauge transformations like (\ref{gauge}) induce
constraints on wave functions
$\psi(U)$. Let us denote by $L^\alpha_\beta$ the \emph{left-invariant} vector fields that generate the
gauge right-translations (\ref{gauge}). They are dual to the
left-invariant 1-forms $\vartheta^L=iU^{-1}dU=\vartheta^\alpha_\beta
X_\alpha^\beta$, so that
$\vartheta^\alpha_\beta(L^\gamma_\sigma)=\delta^\alpha_\sigma\delta^\gamma_\beta$.
For the case of $\mathbb F_{N-1}=G/G^0=U(N)/U(1)^{N}$, constraint
equations attached to the gauge invariance (\ref{gauge}) can be written
as: \be
 L^\alpha_\alpha\psi
 = -2S_\alpha\psi, \;
 \alpha=1,\dots,N,\label{constraint1}
 \ee
where $(S_1,\dots,S_{N})=S$ denote $N$ (half-integer) quantum numbers
(``spin labels'') characterizing the representation. For
complex-projective
$\mathbb CP^{N-1}=U(N)/U(N-1)\times U(1)$ spaces and open complex balls $\mathbb
CD^{N-1}=U(1,N-1)/U(N-1)\times U(1)$, constraint equations read:
\bea L^1_1\psi
 &=& -2S_1\psi,\nn\\
 L_\alpha^\beta\psi&=&0, \ \forall \alpha,\beta>1,\label{constraint3}\eea
where now there is only one quantum number $S_1$ characterizing the
representation. Constraints for other cosets follow similar guidelines.

Moreover, we shall restrict ourselves to the Lagrangian (\ref{Lagrangian})
for coupling constants $\rho_1=\rho$ and $\rho_2=0$ and we shall work in a
holomorphic picture, which means that constrained wave functions
(\ref{constraint1}) will be further restricted by:
\be
 L_\alpha^\beta\psi=0, \ \forall \alpha>\beta=1,\dots,N-1.\label{constraint2}
\ee
Those readers familiar with Geometric Quantization \cite{GQ1,GQ2} will
identify the constraint equations (\ref{constraint1}) and
(\ref{constraint2}) as \emph{polarization} conditions (see also \cite{GAQ}
for a Group Approach to Quantization scheme and \cite{higherpol} for the
extension of first-order polarizations to higher-order polarizations),
intended to reduce the representation of $G$ on wave functions $\psi$ on
$G/G^0$. Also, the constraints (\ref{constraint1}) and
(\ref{constraint2}) are exactly the defining relations of a lowest-weight
representation. Actually, left-invariant vector fields $L_\alpha^\beta$
satisfy (creation and annihilation) harmonic oscillator commutation
relations,
\be [L_\alpha^\beta,L_\beta^\alpha]=2(S_\beta-S_\alpha), \ee
when acting on constrained states (\ref{constraint1}), the constraint
(\ref{constraint2}) then defining a holomorphic (or anti-holomorphic)
representation.

The quantum operators of our theory will be
 the \emph{right-invariant} vector fields $R^\alpha_\beta$ that generate the
left-symmetry translations (\ref{leftrans}). They are dual to the
right-invariant 1-forms $\vartheta^R=idU
U^{-1}=\tilde{\vartheta}^\alpha_\beta X_\alpha^\beta$, so that
$\tilde{\vartheta}^\alpha_\beta(R^\gamma_\sigma)=\delta^\alpha_\sigma\delta^\gamma_\beta$.
We shall continue using the notation $X^\alpha_\beta$, instead of
$R^\alpha_\beta$, for our operators, in the hope that no confusion arises.

Before constructing the corresponding Hilbert space, let us show how to
put coordinates $x^\alpha_\beta$ on $G$.

\subsection{Complex coordinates on $U(N_+,N_-)$}

We shall exemplify our construction with the cases of the compact
$G=U(4)$ and non-compact $G=U(1,3)$ groups, in a unified manner.

In order to put coordinates on
$G$, the ideal choice is the Bruhat decomposition
\cite{Helgason} for the coset space (flag manifold)
$\mathbb F=G/T$, where we denote $T=U(1)^{N}$ the maximal torus.
We shall introduce a local complex parametrization of
$\mathbb F$ by means of the isomorphism
$G/T=G^{\mathbb C}/B$, where
$G^{\mathbb C}\equiv GL(N,\mathbb C)$ is the complexification of $G$, and $B$ is
the Borel subgroup of upper triangular matrices. In one direction, the
element $[U]_T\in G/T$ is mapped to $[U]_B\in G^{\mathbb C}/B$. For
example, for $G=U(4)$ we have:
\be
[U]_T=\bordermatrix{&\u1 &\u2 &\u3 &\u4 \cr &u_{1}^{1}&u^1_2 &u^1_3 &u^1_4
\cr &u^2_1&u^2_2&u^2_3&u^2_4 \cr &u^3_1&u^3_2&u^3_3&u^3_4 \cr
&u^4_1&u^4_2&u^4_3&u^4_4 } \longrightarrow [U]_B=\bordermatrix{&\z1 &\z2
&\z3 &\z4 \cr & 1 & 0&0 &0 \cr & x^2_1 & 1&0&0&\cr& x^3_1&x^3_2&1&0&\cr&
x^4_1&x^4_2&x^4_3&1}\label{triang} \ee
where
\bea & &x^2_1=\frac{u^2_1}{u^1_1},\,\,\,x^3_1=\frac{u^3_1}{u^1_1},\, \,\,
x^4_1=\frac{u^4_1}{u^1_1},\nn\\
&&x^3_2=\frac{u^1_1u^3_2-u^1_2u^3_1}{u^1_1u^2_2-u^1_2u^2_1},\, \,\,
x^4_2=\frac{u^1_1u^4_2-u^1_2u^4_1}{u^1_1u^2_2-u^1_2u^2_1},
\label{su4coord}\\ &&x^4_3=\frac{u^1_3(u^2_1u^4_2-u^2_2u^4_1)-
u^2_3(u^1_1u^4_2-u^1_2u^4_1)
+u^4_3(u^1_1u^2_2-u^1_2u^2_1)}{u^1_3(u^2_1u^3_2-
u^2_2u^3_1)-u^2_3(u^1_1u^3_2-u^1_2u^3_1)
+u^3_3(u^1_1u^2_2-u^1_2u^2_1)},\nn \eea
provides a complex coordinatization $\{x^{\alpha}_{\beta},
\alpha>\beta=1,2,3\}$ of nearly all of the 12-dimensional complex flag
manifold
$\mathbb F_{3}=U(4)/U(1)^4$,
missing only a lower-dimensional subspace; indeed, these coordinates are
defined where the denominators are non-zero.

In the other direction, i.e. from
$G^{\mathbb C}/B$ to $G/T$, one uses the Iwasawa decomposition:
any element $W\in G^{\mathbb C}$ may be factorized as $W=VL, V\in G, L\in
B$ in a unique manner, up to torus elements $t\in T$, the Cartan subgroup
of diagonal matrices $t={\rm diag}(x_{1}^1,x_{2}^2,x_{3}^{3},x_{4}^{4})$,
the coordinates of which
$x_{\alpha}^{\alpha}$ (phases)
 can be calculated in terms of the arguments
of the upper angular minors of order $\alpha$ of $U\in G$ as
 \be x_{1}^{1}=\left(\frac{u^1_1}{\bar{u}_{1}^{1}}\right)^{1/2},\,
x_{2}^{2}=(x_{1}^{1})^{-1}\left(\frac{u^1_1u_{2}^{2}-u^{1}_{2}u_{21}}{\bar{u}^{1}_{1}\bar{u}_{2}^{2}-
\bar{u}^{1}_{2}\bar{u}^{2}_{1}}\right)^{1/2},\label{phases}\dots. \ee
 The Iwasawa decomposition in this case may be proved by means of the
Gram-Schmidt ortonormalization process: regard any $[U]_B\in G^{\mathbb
C}/B$ [like the one in (\ref{triang})] as a juxtaposition of $N$ column
vectors
$(x_1,x_2,\dots,{x}_N)$. Then one obtains orthogonal vectors $\{\w\alpha\}$ in the usual way:
\be
v_\alpha'=\z\alpha- (\z\alpha,v_{\alpha-1}) v_{\alpha-1}-\dots
-(\z\alpha,\w1)\w1,\;\; v_\alpha={v_\alpha'}/{
(\Lambda^{\alpha\alpha}(v_\alpha',v_\alpha'))^{1/2}},\label{g-s} \ee
where
$(x_\alpha,v_\beta)\equiv \bar{x}_{\alpha}^{\mu}\Lambda_{\mu\nu}v_{\beta}^{\nu}$
denotes a scalar product in $\mathbb{C}^N$ with metric $\Lambda$. The
explicit expression of (\ref{g-s}) for the simple cases of the compact
(positive sign) $U(2)$ and non-compact
$U(1,1)$ (negative sign) groups is:
\be v_1^\pm=\frac{1}{\sqrt{\Delta_1^\pm(\bar{x},x)}}\begin{pmatrix}
    1 \\
    {x^{2}_{1}}
  \end{pmatrix}, \;\;  v_2^\pm=\frac{1}{\sqrt{\Delta_1^\pm(\bar{x},x)}}\begin{pmatrix}
     \mp {{x}^{1}_{2}}  \\
    1
  \end{pmatrix},\;\; x^{1}_{2}\equiv \bar{x}^{2}_{1},
\ee
where we define the \emph{length}:
\be
\Delta_1^\pm(\bar{x},x)\equiv 1\pm |x^{2}_{1}|^2. \label{length}\ee
The more involved case of
$U(4)$ can be found in \cite{brokenFD} and that of $U(2,2)$ in \cite{cqg}.
We shall give here only the explicit expressions of the lengths for
$U(4)$ (positive sign) and
$U(1,3)$ (negative sign):
\begin{eqnarray}
\Delta_1^\pm(\bar{x},x)&=&{1\pm|x^{2}_{1}|^2\pm |x^{3}_{1}|^2 \pm
|x^{4}_{1}|^2},\nn
\\ \Delta_2^\pm(\bar{x},x)&=&{1\pm|x^3_2x^{4}_{1}-x^4_2x^3_1|^2+ |x^3_2|^2+
|x^4_2|^2 \pm|x^3_2x^{2}_{1}-x^3_1|^2\pm |x^4_2x^2_1-x^{4}_{1}|^2} \nn\\
\Delta_3^\pm(\bar{x},x)&=&{1+|x^4_3|^2+ |x^4_2-x^4_3x^3_2|^2\pm
|x^{4}_{1}+ x^4_3x^3_2x^2_1-x^4_2x^2_1-x^4_3x^3_1|^2}.\label{lengths}
\end{eqnarray}
These quantities will play an important role in the calculation of
irreducible matrix coefficients (\ref{imc}) through the Bergmann's kernel.
They can be computed as the upper angular minors of order $\alpha=1,2,3$
of the matrix
$\Lambda[U]_B^\dag\Lambda[U]_B$, with $[U]_B$ given in (\ref{triang}) and
$\Lambda={\rm diag}(1,\pm 1,\pm 1,\pm 1)$.

Any unitary matrix $U\in G$ in the present patch (which contains the
identity element, $x^\alpha_\beta=0, \forall \alpha\not=\beta,
x^\alpha_\alpha=1$) can be written in minimal coordinates
$x^\alpha_\beta, \alpha,\beta=1,\dots,N$ (with the definition of $x^\alpha_\beta=\bar{x}^\beta_\alpha, \forall \alpha<\beta$)
as the product $U=Vt$ of an element
$V$ of the base (flag) $\mathbb F_{N-1}$ times an element $t$ of the fibre
$T=U(1)^{N}$. Namely, for $U_+\in U(2)$ and $U_-\in U(1,1)$ we have simply:
\be
U_\pm(x)=Vt=
  \frac{1}{\sqrt{\Delta_1^\pm(x,\bar{x})}}\begin{pmatrix}
    1 &  \mp {{x}^{1}_{2}} \\
    {x^{2}_{1}}& 1
  \end{pmatrix}
  \begin{pmatrix}
     x_{1}^{1} & 0 \\
    0 & x_{2}^{2}
  \end{pmatrix}.\label{matrixel2}
  \ee

\subsection{Unitary irreducible representations and coherent states of $U(N_+,N_-)$\label{highrep}}

We have shown how to compute the matrix elements of a general
(pseudo-)unitary transformation $U(x)\in U(N_+,N_-)$ in the fundamental
representation. Now, let us show how to proceed for more general
(``higher-spin'') unitary irreducible representations. The group
$U(N_+,N_-)$ is non-compact so that, unlike the case of $U(N)$, all its unitary
irreducible representations are infinite-dimensional. Here, we shall
restrict ourselves to representations of the discrete series where each
carrier space ${\cal H}_S$ is labelled by the quantum numbers
(`generalized spin') $S=(S_1,\dots,S_N), S_\alpha\in \mathbb Z/2$ of Eq.
(\ref{constraint1}) and is spanned by an orthonormal basis $B({\cal
H}_S)=\{|S,n\rangle\}$, made of eigenstates of the Cartan subalgebra
(\ref{cartan}), i.e.
\be
X_{\alpha}^{\alpha}|S,n\rangle=(n(\alpha)-2S_\alpha)|S,n\rangle,\label{cartan1}\ee
where the index
$n$ denotes an integral upper-triangular
$N\times N$ matrix with entries $n_{\alpha}^\beta, \alpha>\beta=1,\dots,N-1$ and
\be
n(\alpha)\equiv\sum_{\beta>\alpha}n_{\beta}^\alpha-\sum_{\beta<\alpha}n^{\beta}_\alpha\label{cartan2}\ee
(see \cite{gp} for more details). The range of the entries
$n_{\alpha}^{\beta}$ of $n$ depends on the set of ``spin'' indices
$\{S_\alpha, \alpha=1,\dots,N\}$.

Constrained wave functions (\ref{constraint1},\ref{constraint2}) can be
arranged in general as the product (see \cite{acha} for a general proof):
\be
\psi_s(x)=W_{s}(x)\phi(x^\alpha_\beta),\; \alpha<\beta\label{antiholf} \ee
of an arbitrary holomorphic function $\phi({x}^\alpha_\beta),
\alpha<\beta$ [which is the general solution of Eqs. (\ref{constraint1})
and (\ref{constraint2}) for $S_\alpha=0$], times a lowest-weight vector
$W_s$ (``vacuum''), which is a particular solution of Eqs. (\ref{constraint1})
and (\ref{constraint2}) for $S_\alpha\not=0$. A possible choice of
$W_s$ can be given in terms of lengths like (\ref{lengths}) as (see \cite{gp} for more
details):
\be W_{s}(x)\equiv\prod_{\beta=1}^{N}
\frac{1}{t_\beta^{2{s_\beta}}\Delta_\beta(\bar{x},x)^{s_\beta}},\ee
where we have defined:
\be\Delta_N\equiv 1,\ t_\beta \equiv {x}^{\beta}_{\beta}
\bar{x}^{\beta+1}_{\beta+1}, \ t_N\equiv
\prod_{\beta=1}^{N}{x}^{\beta}_{\beta},\ s_\beta\equiv
S_\beta-S_{\beta+1}, \ s_N\equiv \sum_{\beta=1}^N S_\beta.\ee
The sign of $s_\beta$ is always strictly positive except for
$s_{N_+}\in \mathbb Z^-/2$, in order to ensure the existence of a finite
scalar product (see, e.g., \cite{gp}). In bracket notation we shall write:
\be
W_{s}(x)\equiv\langle s,x|s,0\rangle=\langle
s,0|U^\dag(x)|s,0\rangle,\;\psi_s(x)\equiv\langle s,x|\psi\rangle. \ee
Here we are implicitly making use of the Coherent-States machinery (see
e.g. \cite{Perelomov,Klauder}). Actually, we are denoting by
$|s,x\rangle\equiv U(x)|s,0\rangle$ the set of vectors in the orbit of the
vacuum $|s,0\rangle$ (the lowest-weight vector) under the action of the
group $G$ (this set is called a family of \emph{covariant coherent states}
in the literature \cite{Perelomov,Klauder}). Given that
$|s,n\rangle$  are eigenstates of the Cartan subalgebra (\ref{cartan}),
the Cartan (isotropy) subgroup $T=U(1)^N$ stabilizes the vacuum vector
$|s,0\rangle$ up to multiplicative phase factors (characters of $T$) which are irrelevant for the
calculation of the modulus of the matrix coefficients (\ref{imc}). Thus,
we shall make use of the factorization $U=Vt, V\in \mathbb F_{N-1}, t\in
T$ in (\ref{matrixel2}) and we shall restrict ourselves to classes of
coherent states modulo $T$ (usually referred to as Gilmore-Perelomov
coherent states), that is, we shall set
$t=1$.

Eventually, we are interested in the calculation of the irreducible matrix
coefficients (\ref{imc}). For general pseudo-unitary groups $U(N_+,N_-)$,
this task is quite unwieldy. Fortunately, we are just interested in the
computation of the matrix elements
\be U_{0n}^{(s)}(x)=\langle s,
0|U(x)|s, n\rangle,\label{purpose}\ee
the squared modulus of which gives the average number of particles with
energy
$E_n$ (\ref{averagenp}) in the excited vacuum (\ref{vacp}), up to a
constant factor. For this, we can take advantage of the \emph{coherent
state overlap} or reproducing (normalized Bergmann's) kernel
$\langle s,x|s,x'\rangle=\langle s,0|U(x)^\dag U(x')|s,0\rangle$, which  can  be
expressed in terms of the upper-angular minors of
$\Lambda[U]_B^\dag\Lambda[U]_B$ and $\Lambda[U]_B^\dag\Lambda[U]_B'$ as follows (see e.g. \cite{gp,Perelomov,Klauder}
for more details):
\be
\langle s,x|s,x'\rangle=\prod_{\beta=1}^{N-1}
\frac{\Delta_\beta(\bar{x},x')^{2s_\beta}}{\Delta_\beta(\bar{x},x)^{s_\beta}
\Delta_\beta(\bar{x}',x')^{s_\beta}}.\label{repker}\ee
 For example, for
$U(2)$ and $U(1,1)$ the expression (\ref{repker}) reads:
\be
\langle s,x|s,x'\rangle_\pm=\frac{(1\pm \bar{x}^2_1 x'^2_1)^{\pm
2s}}{(1\pm |x^2_1|^2)^{\pm s} (1\pm |x'^2_1|^2)^{\pm s}},\;\;s=1/2, 1,
3/2, \dots\label{repker2}\ee
where the upper sign corresponds to $U(2)$ and the lower one to
$U(1,1)$. The coherent state overlap (\ref{repker}) will play the role of a
``generating function'' for our matrix elements (\ref{purpose}). In fact,
inserting the resolution of unity $I=\sum_n|s,n\rangle\langle s,n|$ in the
coherent state overlap (\ref{repker}), we arrive at:
\be \langle
s,x|s,x'\rangle=\sum_n\langle s,0|U(x)^\dag|s,n\rangle \langle
s,n|U(x')|s,0\rangle.\ee
We can easily identify the addends of this series
in the expansion of the upper-angular minors $\Delta_\beta(x',\bar{x})$ in
(\ref{repker}). For example, for
$U(2)$ and $U(1,1)$, the expansion of (\ref{repker2}),
\bea
 \langle s,x|s,x'\rangle_+&=&\frac{\sum_{n=0}^{2s}\binom{2s}{n}(\bar{x}^2_1x'^2_1)^n}
 {(1+ |x^2_1|^2)^{s} (1+ |x'^2_1|^2)^{s}}, \nn\\
  \langle s,x|s,x'\rangle_-&=&\frac{\sum_{n=0}^{\infty}\binom{2s+n-1}{n}(\bar{x}^2_1x'^2_1)^n}
 {(1- |x^2_1|^2)^{-s} (1- |x'^2_1|^2)^{-s}},\label{repker3}\eea
gives
\bea U_{0n}^{(s)}(x)_+&=&\langle s, 0|U_+(x)|s, n\rangle=
\frac{\binom{2s}{n}^{1/2}({x}^2_1)^n}
 {(1+ |x^2_1|^2)^{s}},\nn\\ U_{0n}^{(s)}(x)_-&=&\langle s, 0|U_-(x)|s, n\rangle=
\frac{\binom{2s+n-1}{n}^{1/2}({x}^2_1)^n}
 {(1- |x^2_1|^2)^{-s}},\label{matrixukappa}
\eea
for the matrix coefficients (\ref{purpose}). An alternative way of
arriving at these expressions is by exponentiating the usual action
\bea J_0|s,n\rangle&=&(n-\kappa s)|s,n\rangle,\nn\\
J_+|s,n\rangle&=&\sqrt{(n+1)(2s-\kappa n)}|s,n+1\rangle,\nn\\
J_-|s,n\rangle&=&\sqrt{n(2s-\kappa(n-1))}|s,n-1\rangle, \label{repre1}\eea
of the $SU(2)$ ($\kappa=+1$) and $SU(1,1)$ ($\kappa=-1$) generators
$J_+=X^{1}_{2}, J_-=\kappa X^{2}_{1}, J_0=(X^{1}_{1}-X^{2}_{2})/2$, with the standard commutation relations:
\be \left[J_+,J_-\right]=2\kappa J_0, \;\;\left[J_0,J_\pm\right]=\pm
J_\pm. \label{basicalgebra} \ee
Actually, any group element $U(x)\in G_\kappa/T$ can be represented as:
\be
U_\kappa(x)=e^{\zeta J_+-\bar{\zeta}J_-},
\,\,x^{2}_{1}=\frac{\zeta}{|\zeta|}\tan_\kappa|\zeta|,\label{exponential}\ee
where we denote by $\tan_1=\tan$ and $\tan_{-1}=\tanh$, the standard
trigonometric and hyperbolic tangents. Using the expressions
(\ref{repre1}-\ref{exponential}) one can recover again, after some
algebra, the result (\ref{matrixukappa}).

According to the general prescription given in (\ref{hamiltoniancpn}), our
Hamiltonian operator for
$\mathbb CP^1=U(2)/U(1)^2$ and
$\mathbb CB^1=U(1,1)/U(1)^2$ will be:
\be
X_0^\kappa=\hbar\omega(J_0+\kappa s I)=\hbar\omega(
X_{1}^1-\Xi_\kappa),\;\;\omega\equiv\lambda/\hbar,
\label{hamiltonianu2}\ee
where $\Xi_\kappa\equiv X^1_1+X^2_2$ (the zero-point --`dark'-- energy
operator) stands for the linear (trace) Casimir operator fulfilling
\be
\Xi_\kappa|s,n\rangle=-2\kappa s|s,n\rangle.\label{darkenergy}\ee
The reason to subtract the `dark' energy $\Xi_\kappa$ from our Hamiltonian
operator (\ref{hamiltoniancpn}) is in order to fulfil our initial
requirement (\ref{requirement}) that the vacuum state
$|s,0\rangle$ has no energy, $E_0=0$. In fact:
\be
 X_0^\kappa|s,n\rangle=E_n|s,n\rangle,\,E_n=\hbar\omega n\label{spectrum}\ee
[remember also the Eqs. (\ref{cartan1}) and (\ref{cartan2})]. We have
introduced the Planck constant
$\hbar$ and the frequency
$\omega=\lambda/\hbar$ motivated by the spectrum (\ref{spectrum}), which
suggests that we are dealing with a generalized harmonic oscillator;
actually, the standard harmonic oscillator representation is recovered in
(\ref{repre1}) for $\kappa=0$ (zero curvature).

This algorithm can be straightforwardly extended to higher-dimensional
pseudo-unitary groups and higher-dimensional cosets, although the
resulting expressions are much more involved. We shall concentrate
ourselves in the case of coherent states of
$U(N)$ and
$U(1,N-1)$ parametrized by points of the coset spaces $\mathbb CP^{N-1}=U(N)/U(N-1)\times U(1)$ and
$\mathbb CD^{N-1}=U(1,N-1)/U(N-1)\times U(1)$, respectively. This choice
corresponds to wave functions constrained by (\ref{constraint3}), that is,
to picking $s_\alpha=0, \forall \alpha>1$ in (\ref{repker}) and
$s_1=\pm s,\, s\in\mathbb Z^+/2$, for $U(N)$ and $U(1,N-1)$, respectively. More explicitly:
\be
\langle s,x|s,x'\rangle_\pm= \frac{\Delta_1(\bar{x},x')^{\pm
2s}}{\Delta_1(\bar{x},x)^{\pm s} \Delta_1(\bar{x}',x')^{\pm
s}}=\frac{(1\pm\sum_{\alpha=2}^{N}\bar{x}^\alpha_1 x'^\alpha_1)^{\pm
2s}}{(1\pm\sum_{\alpha=2}^{N}|x^\alpha_1|^2)^{\pm
s}(1\pm\sum_{\alpha=2}^{N}|x'^\alpha_1|^2)^{\pm s}} .\label{repker4}\ee
Expanding the numerator of the last fraction, like in (\ref{repker3}), we
arrive at the matrix elements:
\bea U_{0n}^{(s)}(x)_+&=&\langle s, 0|U_+(x)|s, n\rangle=
\frac{\left(\frac{2s!}{\prod_{\alpha=1}^{N}n_\alpha
!}\right)^{1/2}\prod_{\alpha=2}^{N}({x}^{\alpha}_{1})^{n_\alpha}}
 {(1+ \sum_{\alpha=2}^{N}|x_1^{\alpha}|^2)^{s}},\nn\\
 U_{0n}^{(s)}(x)_-&=&\langle s, 0|U_-(x)|s, n\rangle=
\frac{\left(\frac{(2s+n_1-1)!}{(2s-1)!\prod_{\alpha=2}^{N}n_\alpha
!}\right)^{1/2}\prod_{\alpha=2}^{N}({x}^{\alpha}_{1})^{n_\alpha}}
 {(1-\sum_{\alpha=2}^{N}|x_1^{\alpha}|^2)^{-s}},\label{matrixukappa2}
\eea
where now $n=(n_1,\dots,n_{N})$ [or $(n_1^1,\dots,n_{N}^1)$ in the
notation of Eqs. (\ref{cartan1},\ref{cartan2})] is an integral vector
satisfying the constraint
\be
\sum_{\alpha=1}^{N}n_\alpha=2s\label{n1un}\ee
for $U(N)$, and
\be
 \sum_{\alpha=2}^{N}n_\alpha=n_1, \;n_1\in \mathbb N, \label{n1u1n}\ee
for $U(1,N-1)$ (see Ref. \cite{Gitman} for other constructions of coherent
states on $\mathbb CP^{N-1}$ and $\mathbb CD^{N-1}$).

The Hamiltonian operator is again
\be
X_0^\kappa=\hbar\omega(X_{1}^1-\Xi_\kappa), \label{higherhamilcpn1-1}\ee
where now $\Xi_\kappa\equiv\sum_{\alpha=1}^N X^\alpha_\alpha$ (the linear
Casimir operator). The energy spectrum is:
\be
X_0^\kappa|s,n\rangle=E_n|s,n\rangle,\, E_n=\hbar\omega
\sum_{\alpha=2}^{N} n_\alpha.\label{higherhamilcpn1}\ee
This choice fulfils again our prerequisite (\ref{requirement}).

\section{Thermal vacuum radiation: generalized quantum
statistics}\label{genqstat}

Let us show how to associate a thermal bath with the excited vacuum
(\ref{vacp}). We shall restrict ourselves to curved phase spaces of
constant curvature (complex projective spaces $\mathbb CP^{N-1}$ and open
complex balls $\mathbb CB^{N-1}$). We shall compute the mean energy
(\ref{meanenergy}) and show that it matches generalized FD and BE
statistics, depending on the curvature, recovering the traditional cases
for $N=2$.

\subsection{Broken $U(1,1)$ symmetry and BE statistics\label{BEsec}}

The unbroken symmetry generators (\ref{unbrokengen}) for this case are
\[{\cal G}_0=\langle \hat{X}_0, \hat{J}_- \rangle,\]
which generate affine or similitude transformations. The group $G_0$
leaves stable the vacuum (\ref{thetav2}) of the spontaneously broken
theory, that is,
$\hat{X}_0|\theta\rangle=0, \hat{J}_-|\theta\rangle=0$, which (we remind) is a coherent state of zero modes. We
can think of zero modes as \emph{virtual} particles without (``bright'')
energy and as being undetectable (``dark'') by affine observers. However,
we shall show that a general symmetry transformation $U(x)\in U(1,1)$,
which incorporates the broken symmetry generator
$\hat{J}_+$, produces a ``rearrangement'' of the affine vacuum $|\theta\rangle$
and makes it to radiate as a black body.

Taking into account the irreducible matrix coefficients
(\ref{matrixukappa}), the probability $P_n(p,x)$ in Eq. (\ref{probpart})
of observing $p$ particles with energy $E_n$ in
$|\theta'\ra$ is:
\be
P_n(p,x)=\frac{e^{-|\theta|^2}}{p!}\left(
\frac{|\theta|^{2}\binom{2s+n-1}{n}}
 {(1- |x^2_1 |^2)^{-2s}}\right)^p  |{x}^{2}_{1}|^{2np}.
\ee
From (\ref{probest}), we see that the relative probability of observing a
state with total energy
$E_q=\hbar\omega q$ in the excited vacuum $|\theta'\rangle$ can be factored as:
\bea P(E_q)&=&\Omega(E_q)(|x^2_1|^2)^q,\\ \Omega(E_q)&=&
\sum_{\begin{array}{c} p_0,\dots,p_q\in \mathbb N:\\ \sum^q_{n=0} n
p_n=q\end{array}} e^{-(q+1)|\theta|^2}\left((1- |x^2_1
|^2)^{s}|\theta|\right)^{2\sum_{n=0}^qp_n}\prod^q_{n=0}
\frac{\tbinom{2s+n-1}{n}^{p_n}}{p_n!}. \eea
We can associate a {\it thermal bath} with this distribution function by
noticing that $\Omega(E_q)$ behaves as a relative weight proportional to
the number of states with energy
$E_q=\hbar\omega q$; the factor $(|x^2_1 |^2)^q<1$ fits this weight properly to a
temperature $T$ as:
\be
(|x^2_1|^2)^q=e^{q\log|x^2_1
|^2}=e^{-\frac{E_q}{k_BT}},\;\;T\equiv-\frac{\hbar\omega}{k_B\log|x^2_1
|^2}.\ee
The (finite subgroup of the) conformal group in 1+1-dimensions is
$SO(2,2)\simeq SU(1,1)\times SU(1,1)$. Its generators are: Lorentz transformations $M_{\mu\nu}$, space-time translations $P_\mu$,
dilations $D$, and special conformal transformations $K_\mu, \mu=0,1$. We
could identify, for example, $D=J_0$ (dilations),
$P=P_1=J_-$ (spatial translations) and $K=K_1=J_+$ (spatial relativistic uniform
accelerations). Indeed, they reproduce the standard commutation relations
of the conformal group (Lorentz transformations do not appear because we
are restricting ourselves to spatial translations):
\[[K,P]=-2D,\;\; [D,K]=K, \;\;[D,P]=-P.\]
We could think of special conformal transformations $K\in su(1,1)$ as
transitions to a uniformly relativistic accelerated frame (see e.g.
\cite{Hill} and \cite{conforme}), so that $T=\frac{\hbar a} {2\pi ck_B}$
is the temperature associated with a given acceleration
$a\equiv -{2\pi\omega c}/{\log|x^2_1|^2}$.

After some intermediate calculations, the expected value of the total
energy in the accelerated vacuum $|\theta'\rangle$, given by Eq.
(\ref{meanenergy}), proves to be:
\bea {\cal X}_0&=& |\theta|^2(1-|x^2_1|^2)^{2s}\sum_{n=0}^\infty
\hbar\omega n\tbinom{2s+n-1}{n} |x^2_1|^{2n}\nn\\ &=& 2s|\theta|^2
\hbar\omega\frac{|x^2_1 |^2}{1-|x^2_1|^2}=2s|\theta|^2 \frac{\hbar\omega
e^{-\hbar\omega/k_BT}}{1-e^{-\hbar\omega/k_BT}}\,,\label{emediaBE} \eea
which is proportional to the {\it mean energy per mode} of the BE
statistics. In $d$ spatial dimensions, the number of states with frequency
between $\omega$ and $\omega+d\omega$ is proportional to
$\omega^{d-1}$. Thus, for
$d=3$, the spectral distribution of the radiation of the accelerated
vacuum $|\theta'\rangle$ is \emph{Planckian}, i.e. $|\theta'\rangle$
radiates as a \emph{black body}. Still, we could have introduced a
\emph{chemical potential}
$\mu$ by setting $|x^2_1|^2= e^{(\mu-\hbar\omega)/k_BT}$ with the usual
restriction for bosons $\mu<\hbar\omega$, in order to preserve the open
unit ball condition $|x^2_1|<1$.

\subsection{Broken $U(2)$ symmetry and FD statistics\label{FDsec}}

 As for the non-compact case of $U(1,1)$, we can associate again a
thermal bath with the state
$|\theta'\rangle$. The difference now is that the factor $|x^2_1 |$ is unbounded from above,
i.e.
$|x^2_1|<\infty$, which means that we have to introduce a non-zero
\emph{chemical potential} $\mu$, with no restrictions this time, such that
$|x^2_1|^2=e^{(\mu-\hbar\omega)/k_BT}$, or/and to allow for
\emph{negative temperatures}. The expected value of the total energy
(\ref{meanenergy}) in the polarized vacuum
$|\theta'\rangle$ then proves to be:
\bea {\cal X}_0&=&{\langle\theta'|\hat{\ce}|\theta'\rangle}=
|\theta|^2(1+|x^2_1 |^2)^{-2s}\sum_{n=0}^{2s} \hbar\omega n\tbinom{2s}{n}
|x^2_1|^{2n}\nn\\ &=& 2s|\theta|^2
\hbar\omega\frac{|x^2_1|^2}{1+|x^2_1|^2}=2s|\theta|^2
\frac{\hbar\omega}{1+e^{(\hbar\omega-\mu)/k_BT}}\,,\label{emediaFD} \eea
which is proportional to the {\it mean energy per mode} of the FD
statistics.

\subsection{Generalized quantum statistics}

Let us extend the two previous results to higher dimensions, that is,
$N>2$.
Taking into account the matrix elements (\ref{matrixukappa2}) and putting
$|x^{\alpha}_{1}|^2\equiv
e^{(\mu_\alpha-\hbar\omega)/k_BT},\,\alpha=2,\dots,N$, the expression of
the probability distribution (\ref{probest}), for the Hamiltonian given by
(\ref{higherhamilcpn1}), can be again factorized as
$P(E)=\Omega(E) e^{-E/k_BT}$. Moreover, the expectation value of the total
energy (\ref{meanenergy}) in the excited vacuum (\ref{vacp}) can now be
written as:
\bea {\cal X}_0&=& |\theta|^2{\sum}_{n_1,\dots,n_{N}}'
\hbar\omega\sum_{\alpha=2}^{N} n_\alpha|U_{n 0}^{(s)}(x)_\pm|^2\nn
\\ &=&2s|\theta|^2\frac{ \hbar\omega e^{
({\mu_2-\hbar\omega})/{k_BT}}+\dots+ \hbar\omega e^{
({\mu_{N}-\hbar\omega})/{k_BT}}}{1\pm e^{({\mu_2-\hbar\omega})/{k_BT}}\pm
\dots\pm e^{({\mu_{N}-\hbar\omega})/{k_BT}}}, \label{eev} \eea
where $\sum'$ means that the indices $n_\alpha$ are restricted as
described in Eqs. (\ref{n1un}) and (\ref{n1u1n}). One can recognize here a
generalization of the {mean energy per mode} of the FD and BE statistics
for a $N-1$ compound.

Note that the mean energy (\ref{eev}) could also be written in a nice
compact form as:
\be
{\cal X}_0= |\theta|^2 X_0\ln K_B(\bar{x},x),\ee
where
\be
X_0=\hbar\omega \sum_{\alpha=2}^{N}
\bar{x}_1^\alpha\frac{\partial}{\partial \bar{x}_1^\alpha} \ee
and
\be
K_B(\bar{x},x')= \Delta_1(\bar{x},x')^{\pm
2s}=(1\pm\sum_{\alpha=2}^{N}\bar{x}^\alpha_1 x'^\alpha_1)^{\pm 2s} \ee
is the Bergmann's kernel, which coincides with the coherent-state overlap
(\ref{repker4}) up to normalizing factors.

\section{Comments and outlook}

We have given a way of finding out the curvature sign ($\pm$) of our phase
space by looking at the (FD or BE) nature of the thermal vacuum radiation
detected by $G/G_0$ (or, let us say, `accelerated') observers. Unitary
$U(2)$ and pseudo-unitary $U(1,1)$ symmetries are present in many physical
systems and, therefore, we feel that our results are not restrictive but
deep at the core of many fundamental quantum physical theories.

We also feel tempted to take this `curvature-statistics connection' beyond
the scope of this paper and to explore its possible consequences in modern
cosmology. Nowadays everybody accepts that the vacuum (dark) energy
contributes to the cosmological constant, which affects the expansion of
the universe. The vacuum energy density $E_0$ is related to the vacuum
pressure $P_0$ by the thermodynamic equation of state (see e.g.
\cite{vacmith})
\be
E_0=-P_0. \ee
This relation is universal: it does not depend on the particular
microscopic structure of the quantum vacuum. Negative vacuum energies
cause the expansion of empty space to slow down, whereas for $E_0>0$ the
expansion of empty space tends to speed up. Actual observations indicate
that our universe undergoes an accelerated expansion. How is this
expansion related to the curvature of the universe?. Let us see what our
model says about the connection `curvature-vacuum energy'.

We have defined our (first-quantized) Hamiltonian $X_0$ in terms of only
`bright' energy $E_n$. For example, in (\ref{hamiltonianu2}) and
(\ref{higherhamilcpn1-1}) we removed the `dark' (zero-point) energy
$\hbar\omega \Xi_\kappa$ from the original Hamiltonian
$\hbar\omega X^1_1$ to define $X_0$. Hence, the vacuum energy in the broken
second-quantized theory is:
\be E_0^\kappa=-\langle\theta|\hbar\omega
\hat{\Xi}_\kappa|\theta\rangle=2\kappa \hbar\omega s
|\theta|^2,\label{vacur} \ee
 since
$\hat{\Xi}_\kappa$ is proportional to the total number of particles
operator $\hat{N}$ in the second quantized theory, namely
$\hat{\Xi}_\kappa=-2\kappa s\hat{N}$ [remember Eq. (\ref{averagenp})].
Thus, this construction resembles somehow the actual cosmological models,
where hyperbolic spatial geometries [$\kappa=-1$, like the Anti de Sitter
space-time with spatial symmetry $O(2,1)\sim U(1,1)$], have negative
vacuum energy (positive vacuum pressure), whereas for spherical spatial
geometries [$\kappa=1$, like de Sitter space with spatial symmetry
$O(3)\simeq U(2)$], the vacuum pressure is negative and the expansion of empty space
tends to speed up.

Going on with our `curvature-statistics connection' hypothesis in
cosmology, we could in principle discern between hyperbolic and spherical
geometries for our universe by looking at the (BE and FD) nature of the
vacuum thermal radiation seen by `non-inertial' observers. Actually, BE
and FD statistics have different qualitative behaviour at low frequencies.
Before, it would be worth exploring whether this vacuum thermal radiation
exists or not and whether it is observable/detectable by us or it is in
fact already part of the observed cosmic microwave background radiation.

Other advantage of our derivation of the vacuum energy (\ref{vacur}) is
that it is free from the traditional drawbacks of `fine-tuning', also
referred to as the `unbearable lightness of space-time' \cite{vacmith}.
Indeed, we can make the parameter $|\theta|^2$ (the average number of zero
modes in the ground state $|\theta\rangle$) as small as we like, thus
eluding huge vacuum energies for either spherical or hyperbolic
geometries.

Although we have restricted ourselves to curved phase spaces of constant
curvature (complex projective spaces and open complex balls), we also
paved the way for more general coset spaces like, for example, the
`generalized open disk' $\mathbb D=U(2,2)/U(2)\times U(2)$ (which can be
conformally mapped one-to-one to the `tube domain' in the `complex
Minkowski phase space'), where a group-theoretical discussion of the
Fulling-Unruh effect \cite{Fulling,Unruh} is in progress \cite{progress}.

\section*{Acknowledgements}

Work partially supported by the Fundación Séneca, Spanish MICINN and Junta
de Andalucía under projects 08814/PI/08, FIS2008-06078-C03-01/FIS and
FQM219, respectively.


\end{document}